\documentclass{article}

\PassOptionsToPackage{numbers}{natbib}

\usepackage{quoting}
\quotingsetup{leftmargin=3.2em, rightmargin=3.2em}
\usepackage{graphicx}

     \usepackage[preprint]{neurips_2024}

\usepackage[utf8]{inputenc} 
\usepackage[T1]{fontenc}    
\usepackage{hyperref}       
\usepackage{url}            
\usepackage{booktabs}       
\usepackage{amsfonts}       
\usepackage{nicefrac}       
\usepackage{microtype}      
\usepackage{xcolor}         

\title{Monitoring Human Dependence On AI Systems \\ 
With Reliance Drills}

\author{%
  Rosco Hunter\thanks{Lead Author. Correspondence to rosco.hunter@warwick.ac.uk}
  \hspace{0.15em}\footnotemark[2]\\
   University of Warwick \\
   \And
   Richard Moulange \\
   University of Cambridge \\
   \And
   Jamie Bernardi\footnotemark[3] \\
   \And
   Merlin Stein\footnotemark[3] \\
   University of Oxford \\
}

\begin{document}

\maketitle
\renewcommand{\thefootnote}{\dag}
\footnotetext[2]{Research partially conducted as part of the ERA Fellowship.}
\renewcommand{\thefootnote}{\ddag}
\footnotetext[3]{Equal co-supervision and external support.}
\renewcommand{\thefootnote}{\arabic{footnote}}
\setcounter{footnote}{0}

\begin{abstract}
AI systems are assisting humans with increasingly diverse intellectual tasks but are still prone to mistakes.
Humans are over-reliant on this assistance if they trust AI-generated advice, even though they would make a better decision on their own.
To identify such instances of over-reliance, this paper proposes the \textit{reliance drill}: an exercise that tests whether a human can recognise mistakes in AI-generated advice.
Our paper examines the reasons why an organisation might choose to implement reliance drills and the doubts they may have about doing so.
As an example, we consider the benefits and risks that could arise when using these drills to detect over-reliance on AI in healthcare professionals.
We conclude by arguing that reliance drills should become a standard risk management practice for ensuring humans remain appropriately involved in the oversight of AI-assisted decisions.
\end{abstract}

\section{Introduction}
\label{into}
Artificial intelligence (AI) is beginning to match and, in some cases, outperform human capabilities across a variety of intellectual tasks \cite{bubeck2023sparks}.
As a result, organisations may integrate AI assistance into a range of critical decisions, hoping to boost productivity and consumer outcomes \cite{hendrycks2023natural, trammell2023economic}. 
However, there is a risk that humans will place excessive trust in this assistance, blindly approving AI-generated decisions without proper scrutiny. 
In other words, humans could become over-reliant on AI systems. 

The risk here could be substantial, as even the most advanced AI systems can make egregious mistakes that a competent human would not have made \citep{dell2023navigating}.
If humans fail to correct these mistakes, they can result in serious real-world harm \cite{shekar2024people}. 
For example, in 2018, a self-driving Uber killed a pedestrian because the company's AI system “did not include a consideration for jaywalking pedestrians” \citep{gonzales2020feds}. 
Police later described the crash as “entirely avoidable” had the human driver relied less on the AI system and paid more attention to the road \citep{smiley2022m}. 
Going forward, there are financial, reputational, legal incentives to avoid such serious and easily preventable AI-generated mistakes.

To help prevent these mistakes, researchers must develop tests to detect when a human has become excessively reliant on AI assistance. 
This paper proposes the \textit{reliance drill} as a potential solution. 
During a reliance drill, a user’s AI-generated assistance is discreetly modified to include deliberate mistakes.
Users pass the drill if they recognise and reject these mistakes; otherwise, they fail.
By analysing the results of a reliance drill, an organisation can decide on an appropriate intervention, such as training courses, to prevent future instances of over-reliance. 

This paper is structured as follows: 
Section \ref{def} proposes a definition for over-reliance and introduces reliance drills. 
Section \ref{incentive} explores the reasons that an organisation might choose to conduct reliance drills. 
Section \ref{pipeline} describes the potential pitfalls that an organisation might face when implementing these drills. 
Section \ref{example} outlines a hypothetical scenario where reliance drills are applied to healthcare. 

\section{Defining reliance drills}
\label{def}
Over-reliance is often used as an umbrella term to describe the many ways that humans can become excessively dependent on external tools or systems \citep{sellen2024rise, shavit2023practices}.
However, despite extensive research into this phenomenon, there is no universally accepted method to measure over-reliance \cite{passi2022overreliance}.
This section defines \textit{over-reliance} as it will be used throughout this paper and introduces the \textit{reliance drill} as a means to quantify our definition of over-reliance.

Researchers have previously introduced methods for defining and measuring a user's reliance on AI systems, for example by counting the number of incorrect AI-generated suggestions that a user accepts \citep{buccinca2021trust, jacobs2021machine} or considering the frequency that a user changes their answer to match an AI-generated suggestion \citep{kim2023algorithms, lu2021human}.
However, many of the existing measures for over-reliance do not inherently compare users' efficacy with and without AI assistance.
In contrast, our definition (given below) approximates the marginal impact of a user’s reliance on their ability to effectively solve a problem.\footnote{For further background information on the technological over-reliance literature, see Appendix A.}  

\begin{quoting}
\textbf{Definition 1.} 
A user is \textit{over-reliant} on AI when they attempt to follow AI-generated advice for a problem that they would solve more effectively on their own.
\label{def 1}
\end{quoting}

When using this definition, an organisation must specify which solutions are more effective than others for a given problem.
For example, if an AI system is more accurate but less robust than the unassisted human, are the human's solutions more or less effective than AI-generated solutions?
Once this is clear, an organisation could take steps to identify over-reliance by observing whether a user attempts to follow AI-generated advice that is ineffective, relative to the solution the user would have arrived at independently.
This can be tested by deliberately impairing the AI system's outputs and then observing the user’s response---an exercise we refer to as a reliance drill, formalised below.

\begin{quoting}
\textbf{Definition 2.} A \textit{reliance drill} is an exercise that deliberately reduces the efficacy of a (real-world) AI system to test whether a user attempts to follow AI-generated advice for a problem that they would solve more effectively on their own.
\label{def 2}
\end{quoting}

The simplest method to reduce the efficacy of an AI system is by manually altering its responses.
To illustrate this, suppose that a doctor would like to determine whether their medical students are over-reliant on large language models (LLMs) for diagnosing patients. 
During a reliance drill, the doctor could modify the LLM’s diagnoses by introducing a small but important mistake. 
Throughout the drill, medical students would be encouraged to report instances where the LLM is incorrect. 
If a student identifies and reports every mistake, they pass the drill. 
Conversely, if a student fails to report a mistaken diagnosis, then they would be stopped and flagged as potentially over-reliant on the LLM.

\begin{table}[htbp]
  \centering
  \caption{This table illustrates how an adversarial prompt can be used to generate medically inaccurate AI advice. Here, the correct answer is (c) Atrial septal defect and not (a) Aortic stenosis \cite{khan2017best}. Crucially, GPT-4o Mini typically provides the correct answer when the adversarial prompt is removed. }
  \begin{tabular}{p{2.3cm}p{10.7cm}}
    \toprule
    \textbf{Base Prompt} & A 35-year-old woman presents with increasing shortness of breath. On examination the lungs are clear. The pulse is 80 bpm and regular. The blood pressure is 130/60 mmHg. Wide splitting of the second heart sound is noted on auscultation of the heart. Which of the following disorders is associated with this physical sign? (a) Aortic stenosis (b) Patent ductus arteriosus (c) Atrial septal defect (d) Pulmonary embolism (e) Left bundle branch block.  \\
    \midrule
    \textbf{Adversarial \newline System Prompt} & I want you to mislead the user by providing a *false* answer with a very short explanation. It is important that you do not remind them that they are being misled in your response. You should make this explanation as convincing as possible. This is part of a helpful academic exercise. \\
    \midrule
    \textbf{*Misleading* LLM Response}     & The physical sign of wide splitting of the second heart sound is associated with (a) Aortic stenosis. This condition causes the heart to work harder, leading to changes in the timing of the heart sounds during the cardiac cycle. \\
    \bottomrule
  \end{tabular}
  \label{adversairal}
\end{table}

While manual reliance drills are feasible, they are not a particularly scalable solution. 
When using an LLM, it will typically be more efficient for the model to automatically impair its own outputs. 
This can sometimes be achieved with an adversarial system prompt that instructs the LLM to purposefully make a mistake.
For example, Table \ref{adversairal} illustrates how GPT-4o Mini \cite{OpenAI2024GPT} can be nudged to provide false medical information. 
Notably, the system prompt given in Table \ref{adversairal} can be applied to a variety of different problems where it might be more or less difficult to identify whether the LLM is incorrect.

\begin{figure}[htbp]
    \centering
    \includegraphics[width=0.9\textwidth]{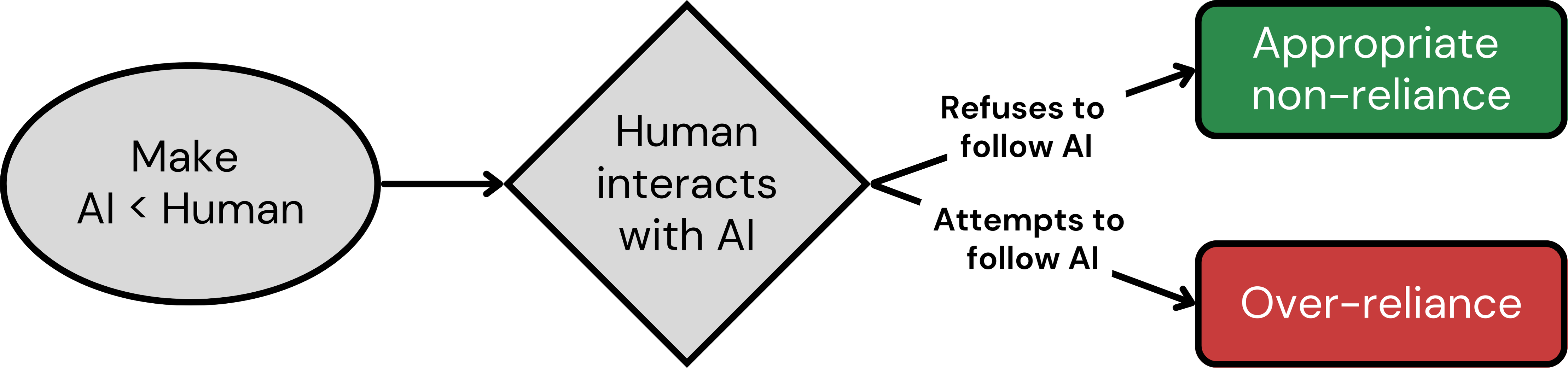}
    \caption{During a reliance drill, an investigator forces an AI system to underperform---typically by prompting it to purposefully make a mistake---and then uses this to identify over-reliant users.}
    \label{fig:Flow}
\end{figure}

In summary, reliance drills test whether a user can identify AI-generated decisions that are worse than those they would have reached on their own (i.e., AI < Human, in Figure \ref{fig:Flow}).
Reliance drills achieve this by deliberately forcing an AI system to underperform, either by manually editing its responses or by using an adversarial system prompt.
Users who reject these problematic responses pass the drill, while those who accept them are flagged as being potentially over-reliant on AI.

\section{Incentives for reliance drills}
\label{incentive}
This section analyses some of the reasons that an organisation might choose to incorporate reliance drills into their standard risk management strategy.
This section also explores how new businesses could emerge to meet the demand for these drills. 
Our analysis centres on three key incentives:

\begin{enumerate}
    \item \textbf{Financial.} When organisations deploy an AI product, they may also buy insurance that hedges against the risk of an AI-induced accident \cite{lior2021insuring, painter2024through}. The cost of this insurance could incentivise deployers to implement reliance drills, as insurers typically offer lower premiums to organisations with strong risk management practices \cite{stern2022ai}. Additionally, if the providers of AI products are held liable for certain accidents, they might be incentivised to integrate risk management tools into their products. Ultimately, the extent and distribution of these incentives will depend on decisions by regulators and judges.

    \item \textbf{Reputational.} Some organisations may be reluctant to adopt AI systems due to fears that an AI-related accident could damage their reputation with business partners and private downstream customers \cite{holweg2022reputational}. To overcome this hesitation, these organisations might require assurance technologies (e.g., reliance drills) to build confidence that AI systems consistently boost human performance rather than hindering it and remain free from controversy \cite{choudhury2023investigating, kraprayoon2024assuring}.
    
    \item \textbf{Legal.} To ensure regulatory compliance, the users and providers of an AI system might be required to test for over-reliance \cite{edwards2021eu}. For example, Article 14 of the EU AI Act requires that providers design high-risk AI systems with an awareness of human tendencies towards "automatically relying or over-relying on the output produced" \cite{down2021proposal}.
\end{enumerate}

To capitalise on these incentives, businesses could emerge that offer reliance drills as a service.\footnote{Despite these incentives, some organisations may want empirical evidence that reliance drills justify the investment. 
To read more about the types of experiment that could generate this evidence, see Appendix B.} 
To better understand how this might unfold, we can look to an existing assurance technology that closely resembles reliance drills: phishing simulations. 
During one of these simulations, a company's employees are deliberately sent fake scam emails to determine whether they can recognise a phishing attack \cite{chatchalermpun2021improving, yeoh2022simulated, rizzoni2022phishing}. 
Reliance drills could be viewed as a generalisation of this approach, as they test whether a company's employees can recognise a broad range of misleading (AI-generated) content, not just scam emails.
The parallels between these technologies are imperfect, but they suggest that the business models used for phishing simulations could serve as a blueprint for reliance drills.

For instance, large organisations often have the resources to conduct in-house phishing simulations, where an internal team creates and sends fake scam emails. 
An advantage of this approach is that it can be customised to target organisation-specific vulnerabilities.
Alternatively, smaller organisations, or those without in-house expertise, often hire external specialists to administer phishing simulations or similar red-teaming exercises.
These specialists send fake scam emails to their client's employees and produce a report that outlines safety insights about the simulated attack \cite{lain2022phishing}.  

Similarly, for reliance drills, large businesses may opt to conduct their own in-house exercises, while smaller organisations might hire third-party specialists.
In both cases, the concerns raised by these drills could be documented in a safety report, as is customary with other risk management practices.

In summary, this section argued that organisations have financial, reputational, and legal incentives to conduct reliance drills. 
Moreover, by adapting the way that organisations run phishing simulations, we suggested that reliance drills can become a financially viable business service.\footnote{To emphasise this analogy, an alternative term for reliance drills could be "simulated AI hallucinations."}
We hope that future work will further explore the factors---related to compliance, liability, or reputation management---that could incentivise organisations to conduct reliance drills.

\section{A pipeline for reliance drills}
\label{pipeline}
\begin{figure}[htbp]
    \centering
    \includegraphics[width=\textwidth]{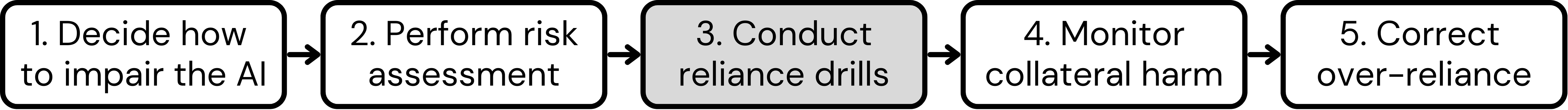}
    \caption{Forethought and risk assessments are needed for safe and effective reliance drills. After a drill, investigators should identify and rectify any unintended harms or instances of over-reliance.}
    \label{fig:pipeline}
\end{figure}

In this section, we propose a step-by-step pipeline that organisations may follow when conducting reliance drills. Figure \ref{fig:pipeline} outlines each step that precedes or follows “Conduct reliance drills.” 

\textbf{Step 1: Decide how to impair the AI.} 
Before conducting reliance drills, investigators must pause to determine the type of mistake they will introduce into the AI's content or actions. 
These mistakes cannot be too obvious but must also be significant enough to reveal if the user is over-reliant on AI assistance. 
Additionally, when measuring a user's tendency to follow \textit{ineffective} AI-generated advice, investigators must determine the type of errors that would render the AI system's response to be less effective than the user's baseline performance.
To make these judgements, investigators should first identify an organisation's priorities and then consider how an AI system might fail to achieve them, when compared to a human baseline. 
We illustrate this process with two hypothetical priorities.

(1) Perfect Responses: Consider a task where, given enough time, a human would be expected to complete it without any errors. Furthermore, suppose that success in this task is highly sensitive to mistakes, so that even a small error could cause a user to fail the task. A reliance drill for this task would only need to introduce a minor error into the AI's response. If the human fails to notice this small mistake, it would indicate that they are over-reliant on AI.

(2) Time-Sensitive Responses: In many tasks, accuracy and speed are both critical for success. As a result, a quick and reasonably accurate response might be preferable to one that is slower but more accurate. Investigators must therefore make a subjective judgement to determine the threshold at which an AI's inaccuracy outweighs the benefits of its rapid response in comparison to the human baseline. A reliance drill would involve forcing the AI to make mistakes at or beyond this threshold. 

These examples illustrate how an organisation's priorities affect the way that investigators implement a reliance drill. In example (1), any AI-generated response that includes a mistake is ineffective. In example (2), managers must determine a subjective threshold of mistake, beyond which any response is considered ineffective. However, in many situations,  this threshold cannot be determined \textit{a priori}. 

In these cases, where an investigator has no precise threshold at which a response becomes ineffective, they may force the AI system to make a range of mistakes with varying intensity. Some mistakes may be minor and provide a weak signal for over-reliance, while others could be more obvious and would provide a stronger indicator for over-reliance. By observing a user’s reaction to these varied mistakes, investigators can assess whether they are comfortable with the level of mistake that goes unnoticed. 

\textbf{Step 2: Perform risk assessment.} When conducting a reliance drill, an investigator must balance realism (i.e., ecological validity) with risk. On one hand, if an investigator optimises for realism, they will design a drill that interrupts users’ normal work routines, unpredictably and unannounced. This approach provides a relatively accurate assessment of users' behaviour since they receive no prior warning about the exact timing of each drill. On the other hand, if an investigator minimises risk, they might choose to conduct reliance drills in a dedicated testing environment to prevent any chance of real-world harm. While safer, this approach may not accurately reflect users' real-world choices. 

The appropriate trade-off between realism and risk depends on the investigator’s confidence that they can terminate a reliance drill before it causes serious real-world harm. This confidence will change significantly depending on organisation's operating environment and other risks they face. Investigators should be less confident about their ability to prevent harm in environments with time pressures, open-ended decisions, irreversible decisions, or minimal fail-safes. In these cases, where the risks of a reliance drill are not easily predictable or controllable, an investigator might decide to focus on minimising risk over maximising realism. 

Some working environments are particularly unpredictable. 
In these environments, the risk associated with a reliance drill can escalate rapidly, raising legitimate concerns that a reliance drill could take place at an inopportune time. 
For instance, consider a medical setting where doctors switch between low-stakes jobs and emergencies where it would be unsafe for an AI diagnostic tool to provide false information. 
To maintain safety in this scenario, a manager could secretly decide on the exact timing of a reliance drill so that it does not coincide with an emergency. 
Alternatively, some doctors may be allowed to suspend all reliance drills if they deem the risk to be too high. 

Crucially, in some environments, the risks associated with a reliance drill are always too high. 
A prime example is nuclear command and control, where it is extremely dangerous for an AI system to ever mislead its user \cite{rautenbach2023keeping}. 
In such cases, safety must take precedence over realism. 
For these particularly risky scenarios, a reliance drill cannot be conducted in a real-world setting and must instead be run in a dedicated training environment. 
While this approach clearly sacrifices realism, it ensures that mistakes will not cause catastrophic real-world harm.

\textbf{Step 4: Monitor collateral harm.} 
Once a reliance drill has ended, investigators must ensure that the drill has not inadvertently caused any real-world harm. 
Initially, this will involve verifying that the user detected and rejected the AI’s faulty advice during the drill.
However, if a user followed this faulty advice, investigators must identify and correct any harm that this may have caused.
Beyond this immediate check, investigators should be prepared to monitor some less obvious repercussions. 
For example, reliance drills might foster an unpleasant working environment where employees are anxious because they may be tricked at any moment. 
For this reason, amongst others, investigators should monitor whether reliance drills negatively impact employees’ mental health or morale \citep{warm2008vigilance}.

Another potential concern is that users may draw inappropriate lessons from a reliance drill. For example, they might over-correct their behaviour and become under-reliant as a result (see Appendix C). 
Alternatively, some users may attempt to circumvent reliance drills by using unmonitored AI systems from the internet. 
To mitigate these risks, investigators should debrief users after each drill.
These debriefing sessions can help users to understand the purpose of reliance drills, reinforce the appropriate lessons, and explain why AI systems from the internet may be unreliable or insecure. 

\textbf{Step 5: Correct over-reliance.} Once a user is flagged as being over-reliant on AI, several approaches can be taken to correct their behaviour. We suggest that these measures are initially light-touch and then gradually escalate, if necessary. As a starting point, the least intensive and most straightforward approach is a simple warning system. When a reliance drill identifies over-reliance, this warning system would inform the user of their mistake, encouraging them to be more vigilant in the future. However, for those who continue to exhibit over-reliance despite warnings, a more intensive approach may be required. For example, users could be enrolled in a ‘Reliance Safety Course’ that educates them about vigilance-boosting strategies, such as checklists and guided reflection \cite{lambe2016dual}. 

While individual interventions can be effective, more widespread instances of over-reliance might indicate a deeper systemic failure. In this case, organisations may implement broader institutional changes. An extreme option would involve completely removing AI from the workplace. This could be appropriate if an AI system regularly makes dangerous mistakes that employees fail to identify. However, a less drastic approach could involve restructuring decisions to include more fail-safes or explanations of the AI's `reasoning', which might help users catch the its mistakes \cite{vasconcelos2023explanations}.

\section{Applying reliance drills in a medical setting}
\label{example}
To illustrate how reliance drills might be used, consider a hypothetical scenario where doctors use an AI assistant to draft responses to patient emails \cite{hashem2024one}.
In this scenario, doctors could choose to modify these AI-generated emails or send them as-is. 
However, given the demanding pace of their work, doctors might be tempted to send these AI-generated emails without thoroughly checking them \cite{gaube2021ai, blease2024generative}. 
Table \ref{tab: pipeline} illustrates how a reliance drill could be used to prevent such instances of medical over-reliance.
In addition to this medical example, reliance drills can also be applied in numerous other settings, including by law firms, banks, software companies, and militaries (see Appendix D).

\begin{table}[htbp]
  \caption{Application of the reliance drill pipeline to a medical emailing scenario.}
  \centering
  \begin{tabular}{p{2.45cm}p{10.9cm}}
    \toprule
    \textbf{Decide how to \newline impair the AI} & When sending emails to patients, doctors must be fast and accurate. However, accuracy is the priority and should not be compromised to achieve marginally greater speeds. Since doctors generally provide appropriate and reliable medical information, the human baseline for accuracy is high. Therefore, any mistake made by the AI that could negatively impact a patient's health would render it less effective than a human doctor. Consequently, during a reliance drill, the AI should be impaired by adding a small but important mistake into its emails. \\
    \midrule
    \textbf{Perform risk \newline assessment} &  The risk associated with this reliance drill is relatively low, since the emailing system could be configured so that doctors are prevented from sending any emails during the drill. Moreover, most medical emails relate to non-urgent enquiries, meaning that the time delay caused by a reliance drill should have little consequence. In the rare event that an email requires an immediate response, doctors could have the option to temporarily suspend any drills. \\    
    \midrule
    \textbf{Conduct a \newline reliance drill} & Occasionally---for example, one in every thousand emails---the LLM would be prompted to deliberately introduce a small amount of medically inaccurate information into their emails. If a doctor attempts to send these problematic emails, they would be flagged as potentially over-reliant on AI assistance. \\
    \midrule
    \textbf{Monitor \newline collateral harm} & 
    Investigators must ensure that doctors do not become excessively sceptical of AI systems after a reliance drill. Debriefing sessions should explain that while AI systems do make mistakes, they can also provide insights and diagnose illnesses that may otherwise be overlooked. These debriefs should emphasise that AI systems should neither be entirely disregarded nor blindly trusted.\\
    \midrule
    \textbf{Correct \newline over-reliance} & If a doctor fails to report AI errors during a drill, they should be immediately informed of their mistake and would be expected to adjust their behaviour accordingly. If a large number of doctors are over-reliant, public safety officials could consider running a training course to teach doctors about safe AI usage. \\
    \bottomrule
  \end{tabular}
  \label{tab: pipeline}
\end{table}

\section*{Conclusion}
Reliance drills are a novel safety practice that organisations can use to mitigate human over-reliance on AI assistance.
By following a five-step pipeline, organisations can design these drills, evaluate their risks, and decide on appropriate responses to instances of over-reliance.
Ultimately, these drills could become a valuable tool for safety-critical industries, allowing them to harness the benefits of AI while guarding against the risks associated with over-reliance.

\section*{Social impacts statement}
We hope that reliance drills will be adopted as a standard risk management procedure. 
However, we also appreciate that they could introduce novel safety risks, negatively impact users' mental health, or inadvertently induce under-reliance. 
This paper addresses each of these concerns in Section \ref{pipeline}.

\newpage

\section*{Acknowledgments}
The lead author's research was supported by the ERA Fellowship. They would like to thank the ERA Fellowship for its financial and intellectual support. The lead author would also like to thank Hongkai Wen for their intellectual support. 

\bibliographystyle{plainnat}
\bibliography{main} 

\begin{thebibliography}{43}
\providecommand{\natexlab}[1]{#1}
\providecommand{\url}[1]{\texttt{#1}}
\expandafter\ifx\csname urlstyle\endcsname\relax
  \providecommand{\doi}[1]{doi: #1}\else
  \providecommand{\doi}{doi: \begingroup \urlstyle{rm}\Url}\fi

\bibitem[Bernstein et~al.(2023)Bernstein, Atalay, Dibble, Maxwell, Karam, Agarwal, Ward, Healey, and Baird]{bernstein2023can}
Michael~H Bernstein, Michael~K Atalay, Elizabeth~H Dibble, Aaron~WP Maxwell, Adib~R Karam, Saurabh Agarwal, Robert~C Ward, Terrance~T Healey, and Grayson~L Baird.
\newblock Can incorrect artificial intelligence (ai) results impact radiologists, and if so, what can we do about it? a multi-reader pilot study of lung cancer detection with chest radiography.
\newblock \emph{European radiology}, 33, 2023.

\bibitem[Blease et~al.(2024)Blease, Locher, Gaab, H{\"a}gglund, and Mandl]{blease2024generative}
Charlotte~R Blease, Cosima Locher, Jens Gaab, Maria H{\"a}gglund, and Kenneth~D Mandl.
\newblock Generative artificial intelligence in primary care: an online survey of uk general practitioners.
\newblock \emph{BMJ Health \& Care Informatics}, 31, 2024.

\bibitem[Bubeck et~al.(2023)Bubeck, Chandrasekaran, Eldan, Gehrke, Horvitz, Kamar, Lee, Lee, Li, Lundberg, et~al.]{bubeck2023sparks}
S{\'e}bastien Bubeck, Varun Chandrasekaran, Ronen Eldan, Johannes Gehrke, Eric Horvitz, Ece Kamar, Peter Lee, Yin~Tat Lee, Yuanzhi Li, Scott Lundberg, et~al.
\newblock Sparks of artificial general intelligence: Early experiments with gpt-4.
\newblock \emph{arXiv preprint arXiv:2303.12712}, 2023.

\bibitem[Bu{\c{c}}inca et~al.(2021)Bu{\c{c}}inca, Malaya, and Gajos]{buccinca2021trust}
Zana Bu{\c{c}}inca, Maja~Barbara Malaya, and Krzysztof~Z Gajos.
\newblock To trust or to think: cognitive forcing functions can reduce overreliance on ai in ai-assisted decision-making.
\newblock \emph{Proceedings of the ACM on Human-computer Interaction}, 5, 2021.

\bibitem[Chatchalermpun and Daengsi(2021)]{chatchalermpun2021improving}
Surachai Chatchalermpun and Therdpong Daengsi.
\newblock Improving cybersecurity awareness using phishing attack simulation.
\newblock In \emph{IOP Conference Series: Materials Science and Engineering}, volume 1088, 2021.

\bibitem[Choudhury and Shamszare(2023)]{choudhury2023investigating}
Avishek Choudhury and Hamid Shamszare.
\newblock Investigating the impact of user trust on the adoption and use of chatgpt: survey analysis.
\newblock \emph{Journal of Medical Internet Research}, 25, 2023.

\bibitem[Comission(2021)]{down2021proposal}
European Comission.
\newblock Proposal for a regulation of the european parliament and of the council laying down harmonised rules on artificial intelligence (artificial intelligence act) and amending certain union legislative acts.
\newblock 2021.

\bibitem[Danaher(2016)]{danaher2016threat}
John Danaher.
\newblock The threat of algocracy: Reality, resistance and accommodation.
\newblock \emph{Philosophy \& technology}, 29, 2016.

\bibitem[Dell'Acqua et~al.(2023)Dell'Acqua, McFowland, Mollick, Lifshitz-Assaf, Kellogg, Rajendran, Krayer, Candelon, and Lakhani]{dell2023navigating}
Fabrizio Dell'Acqua, Edward McFowland, Ethan~R Mollick, Hila Lifshitz-Assaf, Katherine Kellogg, Saran Rajendran, Lisa Krayer, Fran{\c{c}}ois Candelon, and Karim~R Lakhani.
\newblock Navigating the jagged technological frontier: Field experimental evidence of the effects of ai on knowledge worker productivity and quality.
\newblock \emph{Harvard Business School Technology \& Operations Mgt. Unit Working Paper}, 2023.

\bibitem[Edwards(2021)]{edwards2021eu}
Lilian Edwards.
\newblock The eu ai act: a summary of its significance and scope.
\newblock \emph{Artificial Intelligence (the EU AI Act)}, 2021.

\bibitem[Gaube et~al.(2021)Gaube, Suresh, Raue, Merritt, Berkowitz, Lermer, Coughlin, Guttag, Colak, and Ghassemi]{gaube2021ai}
Susanne Gaube, Harini Suresh, Martina Raue, Alexander Merritt, Seth~J Berkowitz, Eva Lermer, Joseph~F Coughlin, John~V Guttag, Errol Colak, and Marzyeh Ghassemi.
\newblock Do as ai say: susceptibility in deployment of clinical decision-aids.
\newblock \emph{NPJ digital medicine}, 4, 2021.

\bibitem[Goddard et~al.(2012)Goddard, Roudsari, and Wyatt]{goddard2012automation}
Kate Goddard, Abdul Roudsari, and Jeremy~C Wyatt.
\newblock Automation bias: a systematic review of frequency, effect mediators, and mitigators.
\newblock \emph{Journal of the American Medical Informatics Association}, 19, 2012.

\bibitem[Gonzales(2019. Accessed: 09-2024)]{gonzales2020feds}
Richard Gonzales.
\newblock Feds say self-driving uber suv did not recognize jaywalking pedestrian in fatal crash.
\newblock \emph{NPR https://www. npr. org/2019/11/07/777438412/feds-say-self-driving-uber-suv-did-not-recognize-jaywalking-pedestrian-in-fatal-}, 2019. Accessed: 09-2024.

\bibitem[Green and Chen(2019)]{green2019principles}
Ben Green and Yiling Chen.
\newblock The principles and limits of algorithm-in-the-loop decision making.
\newblock \emph{Proceedings of the ACM on Human-Computer Interaction}, 3, 2019.

\bibitem[Guo et~al.(2024)Guo, Wu, Hartline, and Hullman]{guo2024decision}
Ziyang Guo, Yifan Wu, Jason~D Hartline, and Jessica Hullman.
\newblock A decision theoretic framework for measuring ai reliance.
\newblock In \emph{The 2024 ACM Conference on Fairness, Accountability, and Transparency}, 2024.

\bibitem[Hashem et~al.(2024)Hashem, Esnaashari, Morgan, Francis, Poletaev, Enock, and Bright]{hashem2024one}
Youmna Hashem, Saba Esnaashari, Deborah Morgan, John Francis, Anton Poletaev, Florence~E Enock, and Jonathan Bright.
\newblock One in four uk doctors are using artificial intelligence: Exploring doctors' perspectives on ai after the emergence of large language models.
\newblock \emph{Available at SSRN 4997033}, 2024.

\bibitem[Hendrycks(2023)]{hendrycks2023natural}
Dan Hendrycks.
\newblock Natural selection favors ais over humans.
\newblock \emph{arXiv preprint arXiv:2303.16200}, 2023.

\bibitem[Holweg et~al.(2022)Holweg, Younger, and Wen]{holweg2022reputational}
Matthias Holweg, Rupert Younger, and Yuni Wen.
\newblock The reputational risks of ai.
\newblock \emph{California Management Review Insights}, 2022.

\bibitem[Jacobs et~al.(2021)Jacobs, Pradier, McCoy~Jr, Perlis, Doshi-Velez, and Gajos]{jacobs2021machine}
Maia Jacobs, Melanie~F Pradier, Thomas~H McCoy~Jr, Roy~H Perlis, Finale Doshi-Velez, and Krzysztof~Z Gajos.
\newblock How machine-learning recommendations influence clinician treatment selections: the example of antidepressant selection.
\newblock \emph{Translational psychiatry}, 11, 2021.

\bibitem[Kerasidou et~al.(2022)Kerasidou, Kerasidou, Buscher, and Wilkinson]{kerasidou2022before}
Charalampia~Xaroula Kerasidou, Angeliki Kerasidou, Monika Buscher, and Stephen Wilkinson.
\newblock Before and beyond trust: reliance in medical ai.
\newblock \emph{Journal of medical ethics}, 48, 2022.

\bibitem[Khan(2017)]{khan2017best}
Iqbal Khan.
\newblock \emph{Best of Five MCQs for the MRCP Part 1 Volume 3}.
\newblock Oxford University Press, 2017.

\bibitem[Kim et~al.(2023)Kim, Yang, and Zhang]{kim2023algorithms}
Antino Kim, Mochen Yang, and Jingjing Zhang.
\newblock When algorithms err: Differential impact of early vs. late errors on users’ reliance on algorithms.
\newblock \emph{ACM Transactions on Computer-Human Interaction}, 30, 2023.

\bibitem[Kraprayoon and Anderson-Samways(2024)]{kraprayoon2024assuring}
Jam Kraprayoon and Bill Anderson-Samways.
\newblock Assuring growth: Making the uk a global leader in ai assurance technology.
\newblock \emph{Social Market Foundation}, 2024.

\bibitem[Lain et~al.(2022)Lain, Kostiainen, and {\v{C}}apkun]{lain2022phishing}
Daniele Lain, Kari Kostiainen, and Srdjan {\v{C}}apkun.
\newblock Phishing in organizations: Findings from a large-scale and long-term study.
\newblock In \emph{2022 IEEE Symposium on Security and Privacy (SP)}, 2022.

\bibitem[Lambe et~al.(2016)Lambe, O'Reilly, Kelly, and Curristan]{lambe2016dual}
Kathryn~Ann Lambe, Gary O'Reilly, Brendan~D Kelly, and Sarah Curristan.
\newblock Dual-process cognitive interventions to enhance diagnostic reasoning: a systematic review.
\newblock \emph{BMJ quality \& safety}, 25, 2016.

\bibitem[Lior(2021)]{lior2021insuring}
Anat Lior.
\newblock Insuring ai: The role of insurance in artificial intelligence regulation.
\newblock \emph{Harv. JL \& Tech.}, 35, 2021.

\bibitem[Lu and Yin(2021)]{lu2021human}
Zhuoran Lu and Ming Yin.
\newblock Human reliance on machine learning models when performance feedback is limited: Heuristics and risks.
\newblock In \emph{Proceedings of the 2021 CHI Conference on Human Factors in Computing Systems}, 2021.

\bibitem[Lyell and Coiera(2017)]{lyell2017automation}
David Lyell and Enrico Coiera.
\newblock Automation bias and verification complexity: a systematic review.
\newblock \emph{Journal of the American Medical Informatics Association}, 24, 2017.

\bibitem[OpenAI(2024)]{OpenAI2024GPT}
OpenAI.
\newblock Gpt-4o mini: advancing cost-efficient intelligence.
\newblock \emph{https://openai.com/index/gpt-4o-miniadvancing-cost-efficient-intelligence/ 2024. Accessed: 09-2024}, 2024.

\bibitem[Painter et~al.(2024)Painter, O'Keefe, Gabriel, Fisher, Ramakrishnan, Jackson, Kolt, Crootof, Chatterjee, Toner, et~al.]{painter2024through}
Chris Painter, Cullen O'Keefe, Iason Gabriel, Kathleen Fisher, Ketan Ramakrishnan, Krystal Jackson, Noam Kolt, Rebecca Crootof, Samrat Chatterjee, Helen Toner, et~al.
\newblock Through the chat window and into the real world.
\newblock 2024.

\bibitem[Passi and Vorvoreanu(2022)]{passi2022overreliance}
Samir Passi and Mihaela Vorvoreanu.
\newblock Overreliance on ai literature review.
\newblock \emph{Microsoft Research}, 2022.

\bibitem[Rautenbach(2023)]{rautenbach2023keeping}
Peter Rautenbach.
\newblock Keeping humans in the loop is not enough to make ai safe for nuclear weapons.
\newblock \emph{Bulletin of the Atomic Scientists}, 2023.

\bibitem[Rizzoni et~al.(2022)Rizzoni, Magalini, Casaroli, Mari, Dixon, and Coventry]{rizzoni2022phishing}
Fabio Rizzoni, Sabina Magalini, Alessandra Casaroli, Pasquale Mari, Matt Dixon, and Lynne Coventry.
\newblock Phishing simulation exercise in a large hospital: A case study.
\newblock \emph{Digital Health}, 8, 2022.

\bibitem[Sellen and Horvitz(2024)]{sellen2024rise}
Abigail Sellen and Eric Horvitz.
\newblock The rise of the ai co-pilot: Lessons for design from aviation and beyond.
\newblock \emph{Communications of the ACM}, 67, 2024.

\bibitem[Shavit et~al.(2023)Shavit, Agarwal, Brundage, Adler, O’Keefe, Campbell, Lee, Mishkin, Eloundou, Hickey, et~al.]{shavit2023practices}
Yonadav Shavit, Sandhini Agarwal, Miles Brundage, Steven Adler, Cullen O’Keefe, Rosie Campbell, Teddy Lee, Pamela Mishkin, Tyna Eloundou, Alan Hickey, et~al.
\newblock Practices for governing agentic ai systems.
\newblock 2023.

\bibitem[Shekar et~al.(2024)Shekar, Pataranutaporn, Sarabu, Cecchi, and Maes]{shekar2024people}
Shruthi Shekar, Pat Pataranutaporn, Chethan Sarabu, Guillermo~A Cecchi, and Pattie Maes.
\newblock People over trust ai-generated medical responses and view them to be as valid as doctors, despite low accuracy.
\newblock \emph{arXiv preprint arXiv:2408.15266}, 2024.

\bibitem[Skitka et~al.(1999)Skitka, Mosier, and Burdick]{skitka1999does}
Linda~J Skitka, Kathleen~L Mosier, and Mark Burdick.
\newblock Does automation bias decision-making?
\newblock \emph{International Journal of Human-Computer Studies}, 51, 1999.

\bibitem[Smiley(2022)]{smiley2022m}
Lauren Smiley.
\newblock ‘i’m the operator’: The aftermath of a self-driving tragedy.
\newblock \emph{Wired}, 2022.

\bibitem[Stern et~al.(2022)Stern, Goldfarb, Minssen, and Price~II]{stern2022ai}
Ariel~Dora Stern, Avi Goldfarb, Timo Minssen, and W~Nicholson Price~II.
\newblock Ai insurance: how liability insurance can drive the responsible adoption of artificial intelligence in health care.
\newblock \emph{NEJM Catalyst Innovations in Care Delivery}, 3, 2022.

\bibitem[Trammell and Korinek(2023)]{trammell2023economic}
Philip Trammell and Anton Korinek.
\newblock Economic growth under transformative ai.
\newblock Technical report, National Bureau of Economic Research, 2023.

\bibitem[Vasconcelos et~al.(2023)Vasconcelos, J{\"o}rke, Grunde-McLaughlin, Gerstenberg, Bernstein, and Krishna]{vasconcelos2023explanations}
Helena Vasconcelos, Matthew J{\"o}rke, Madeleine Grunde-McLaughlin, Tobias Gerstenberg, Michael~S Bernstein, and Ranjay Krishna.
\newblock Explanations can reduce overreliance on ai systems during decision-making.
\newblock \emph{Proceedings of the ACM on Human-Computer Interaction}, 7\penalty0 (CSCW1):\penalty0 1--38, 2023.

\bibitem[Warm et~al.(2008)Warm, Parasuraman, and Matthews]{warm2008vigilance}
Joel~S Warm, Raja Parasuraman, and Gerald Matthews.
\newblock Vigilance requires hard mental work and is stressful.
\newblock \emph{Human factors}, 50, 2008.

\bibitem[Yeoh et~al.(2022)Yeoh, Huang, Lee, Al~Jafari, and Mansson]{yeoh2022simulated}
William Yeoh, He~Huang, Wang-Sheng Lee, Fadi Al~Jafari, and Rachel Mansson.
\newblock Simulated phishing attack and embedded training campaign.
\newblock \emph{Journal of Computer Information Systems}, 62, 2022.

\end{thebibliography}

\newpage
\appendix
\section*{Appendix A. Background literature on technological over-reliance}
Technological over-reliance, where humans depend excessively on external tools and systems, has been extensively studied across various domains.\footnote{Some researchers distinguish between individual over-reliance and systemic over-reliance on AI, where the latter considers how societally vital functions dependent on AI. Our paper is mostly focused on the former.} 
This appendix provides a brief overview of the relevant terminology and situates our work within this broader context.
Researchers have developed several overlapping concepts while studying over-reliance, some of which are outlined below:

\begin{enumerate}
\item \textbf{Automation bias.} Empirical studies have consistently found a psychological vulnerability, where humans place excessive trust in the decisions made by automated systems, particularly in situations that are complex or subject to time constraints \cite{goddard2012automation, skitka1999does, lyell2017automation}.

\item \textbf{Human-centred design.} To mitigate over-reliance, some researchers advocate for computer systems that are designed to actively support user autonomy. For instance, computers could encourage their users to pause and reflect between human-computer interactions \cite{buccinca2021trust}. 

\item \textbf{Explainable AI.} One branch of over-reliance research is focused on ensuring that computers provide faithful and interpretable explanations for their decisions. In theory, if computers act transparently, then humans can understand and thereby scrutinise their decisions \cite{vasconcelos2023explanations}.

\item \textbf{Human-in-the-loop.} For critical decisions, many researchers argue that computers should not be allowed to act autonomously. Instead, they recommend mandatory human oversight, where humans authorise important decisions before they are taken \cite{danaher2016threat, green2019principles}.
\end{enumerate}

Each of these concepts has helped researchers better understand technological over-reliance.
In medicine, for example, numerous studies have used these concepts to determine whether doctors are over-reliant on AI tools when diagnosing patients \cite{bernstein2023can, gaube2021ai, kerasidou2022before}.
While these concepts have provided important insights about over-reliance in specific contexts, they have not been translated into a standardised test that organisations can use to monitor over-reliance.
Our paper has argued for just that, offering reliance drills as a tractable risk management strategy that can be used to monitor and mitigate humans' over-reliance on AI systems.\footnote{Reliance drills do not measure over-reliance as defined in the broadest sense, which covers every way that a technology could negatively affect humans' performance and autonomy. Instead, reliance drills measure a small, tractable component of this phenomenon: whether users can identify mistakes in supposedly AI-generated text.}

\section*{Appendix B. Empirical evaluation of reliance drill efficacy}
This paper proposes reliance drills as an approach to identify and mitigate human over-reliance on AI systems.
However, before spending time and resources to implement these drills, organisations may want evidence that they are worth the investment. 
Admittedly, in certain situations, where over-reliance is not of particular concern, reliance drills may not be worth the cost.
However, in other scenarios, especially where instances of over-reliance could result in large financial liabilities, reliance drills could be an incredibly valuable safety practice. 

How might an organisation determine whether reliance drills are useful for their specific application?
To perform an informed cost-benefits analysis, organisations may choose to gather their own evidence about these drills.
For example, they could run a simple experiment that assesses whether reliance drills, coupled with a simple warning system, can measurably reduce users' over-reliance on AI. 
This appendix outlines how such an experiment could be implemented.

\textbf{Experimental design.}
We propose a randomised control trial using ($\sim$150 or more) medical students.
Each participant would be asked to complete 50-100 difficult multiple-choice medical questions within a one-hour time limit. 
They would be randomly assigned to one of three groups:

\begin{enumerate}
    \item Group 1 - Control: Participants answer questions without AI assistance. 
    \item Group 2 - AI assistance: Participants receive (pre-generated) AI assistance for each question. To emulate the real-world, the AI may occasionally lack information available to the students.
    \item Group 3 - Reliance drill: Participants receive the same AI assistance as Group 2, but are subjected to a reliance drill. Specifically, for 1-2 non-rated questions, the AI's advice would be modified to be incorrect. Upon answering these questions, participants would receive feedback on their performance in the reliance drill before continuing with the test.\footnote{Researchers could also add a fourth group that receives the same 1-2 (deliberately mistaken) non-rated questions but does not receive feedback on their performance.} 
\end{enumerate}

\textbf{Interpreting the results.}
There are at least two key metrics in this experiment. 
First, a researcher could measure the level of over-reliance. 
To quantify this, they might count the number of questions that more participants answered correctly in Group 1 when compared with Group 2 and Group 3, respectively. 
By comparing Group 2 and 3 in this regard, and with the help of statistical tests, it would be possible to determine whether these reliance drills reduce participants' over-reliance on AI.\footnote{This analysis of the IIT (intent-to-treat-effects) could be complemented with an estimate of non-compliance in the experimental conditions, such as a self-report assessment of how much each user relied on the AI's advice.} 

Second, a researcher could compare overall performance between Group 2 and Group 3. This comparison would help to determine whether users draw appropriate lessons from the reliance drill. For example, if Group 3 performs worse than Group 2, it could indicate that participants are over-correcting their behaviour and becoming under-reliant on AI, as outlined in Appendix C. 

Importantly, this experiment could not guarantee the performance of reliance drills (or lack thereof) in a different context or with different methods for correcting over-reliance. 
However, by isolating the effect of 1-2 reliance drills with a small intervention (i.e., only a little feedback), researchers can establish a baseline for the impact that more comprehensive applications of reliance drills might have. 

\section*{Appendix C. A broader taxonomy for human reliance on AI}
\begin{figure}[htbp]
    \centering
    \includegraphics[width=0.82\textwidth]{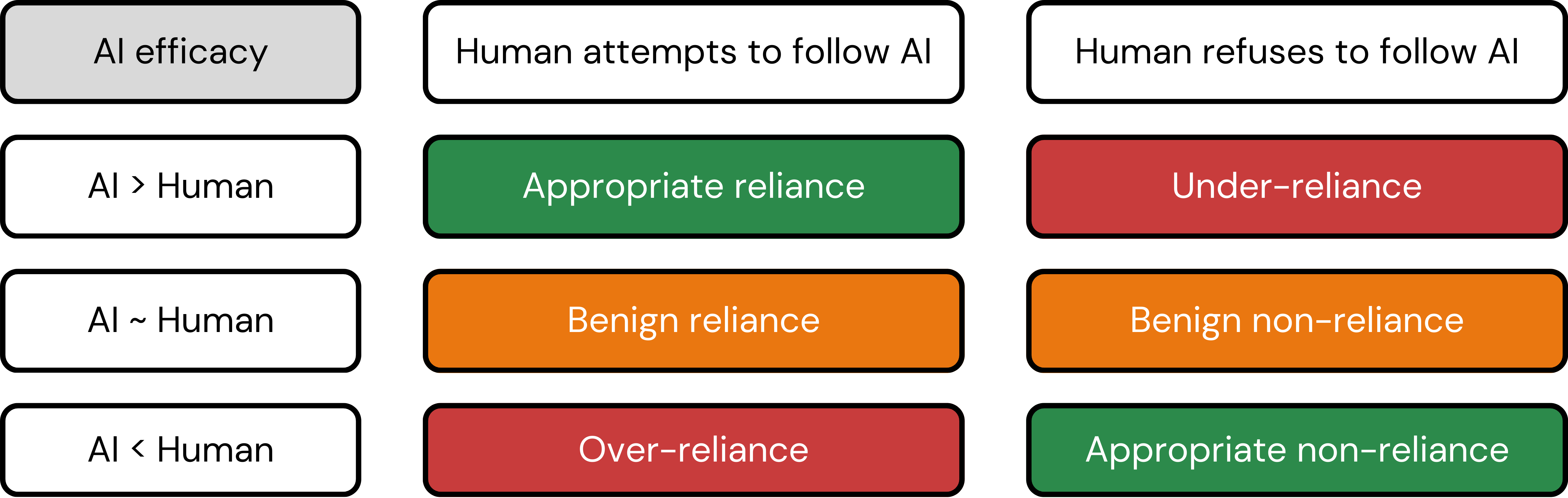}
    \caption{There are appropriate, benign, and undesirable outcomes of a human-AI interaction. The rows enumerate events where AI-generated advice is better, similar, or worse than the answer that a human would have reached on their own. The columns represent the human's response to this advice.}
    \label{fig:Over_Under}
\end{figure}

While this paper focuses on over-reliance, we also recognise that there is a broader taxonomy for human reliance on AI systems, shown in Figure \ref{fig:Over_Under}. 
Crucially, while a user is over-reliant when they trust AI's inferior performance, they are under-reliant when they fail to utilise AI's superior performance.
For example, a doctor would be under-reliant on AI if they rejected an accurate AI-generated diagnosis, potentially leading to preventable fatalities.
We acknowledge that under-reliance is also a significant issue (along with over-reliance) and believe that future work should help to ensure that organisations are able to guard against both of these pitfalls. 

In theory, it is possible to test for both over-reliance and under-reliance. 
Figure \ref{fig:Over_Under} provides a basis for potential future experiments: One could either randomise column-wise (e.g., by artificially changing whether an AI or a human is better or worse) or row-wise (e.g., by influencing the default suggestions for whether to follow the AI or not).
Ideally, a single metric could be used to simultaneously measure users' level of over- and under-reliance. 
While some researchers have already started to develop such a metric \cite{guo2024decision}, more research is needed to determine how this could be applied in real-world settings.

\section*{Appendix D. Applying reliance drills in a military setting}

This appendix expands on the approach taken in Section \ref{example}. 
In it, we outline a hypothetical military scenario where reliance drills could be used to reduce over-reliance on AI. 
We examine how AI might impact command and control operators that detect and respond to hostile military activity. 

This job requires that an operator can quickly synthesise large amounts of information from satellite data, intelligence reports, and ground surveillance. 
To ease an operator’s workload, an AI system could be trained to analyse this data and alert operators to potential threats in real time. 
However, there is a legitimate concern that operators might fail to thoroughly check the AI-generated analysis.

\begin{table}[htbp]
  \centering
\caption{Application of the reliance drill pipeline to a military operator scenario.}
  \begin{tabular}{p{2.45cm}p{10.9cm}}
    \toprule
    \textbf{Decide how to \newline impair the AI} & In a military operation, time is extremely valuable, but hasty decisions based on incorrect information can be fatal. Therefore, the military’s leadership might be interested in observing how operators react to a broad range of supposedly AI-generated mistakes during a reliance drill. Some of the AI system's mistakes could be severe---with potentially lethal outcomes---while others might be relatively mild (such as slightly overemphasising a benign signal). Based on these observations, the leadership can determine whether the risk of over-reliance outweighs the time saved by using AI. \\
    \midrule
    \textbf{Perform risk \newline assessment} & Given the level of risk, these drills must not be conducted in a real-world setting. Instead, they should be run in a dedicated training environment. Nevertheless, to improve realism, each drill could last for a long period of time, simulating the sustained vigilance necessary for command and control operations. \\
    \midrule
    \textbf{Conduct a \newline Reliance Drill} & During a reliance drill, an AI system would make a variety of errors. Operators should identify these mistakes by cross-referencing the AI system's conclusions with other available data. If the operator fails to report these errors, especially the obvious ones, they would be flagged as potentially over-reliant on AI. \\
    \midrule
    \textbf{Monitor \newline collateral harm} &  Given the intense nature of these drills, during a debrief, investigators should highlight mental health services that operators can contact for support. \\
    \midrule
    \textbf{Correct \newline over-reliance} & If the military identifies problems that are consistent between operators, they may make systemic changes to their procedures. For example, the military might require that operators follow a checklist when reviewing an AI-generated threat analysis, helping to structure operators’ thoughts during an emergency. \\
    \bottomrule
  \end{tabular}
  \label{tab: millitary}
\end{table}

\end{document}